\documentclass[sigconf,screen,natbib=false,nonacm]{acmart}

\AtBeginDocument{}

\RequirePackage[
  datamodel=acmdatamodel,
  style=acmnumeric,
  ]{biblatex}

\usepackage{tabularx}
\usepackage{csquotes}
\usepackage{soul}
\usepackage{xspace}
\usepackage{enumitem}
\usepackage{lineno}
\usepackage{tikz}
\usetikzlibrary{arrows,decorations.pathmorphing,backgrounds,calc,positioning,fit,petri,matrix,backgrounds, decorations.pathreplacing, shapes.geometric}
\usepackage{xifthen}
\usepackage{tcolorbox}
\tcbuselibrary{skins,breakable,xparse,raster}
\usepackage{subcaption}
\usepackage{fvcols}
\usepackage{adjustbox}

\newcommand*{\Iltis}{\anonymize{\textsc{Iltis}\xspace}{\anon\xspace}}

\renewcommand{\anon}{$\blacksquare\blacksquare\blacksquare\blacksquare\blacksquare$}
\newcommand{\anonshort}{$\blacksquare\blacksquare\blacksquare$}
\newcommand{\anonymize}[2]{#1}

\newcommand{\algorithmicProblem}[1]{\textnormal{\textsc{#1}}\xspace}

\colorlet{newNodeEdge}{blue!50}
\colorlet{newNodeEdgeText}{black!15!newNodeEdge}
\colorlet{oldNodeEdge}{gray}
\colorlet{oldNodeEdgeText}{black!20!oldNodeEdge}
\colorlet{globalNode}{tugreen}
\definecolor{solutionCol}{HTML}{f49016}
\tikzstyle{solutionNode} = [draw = solutionCol, very thick]

\tikzstyle{originnode} = [draw, circle, inner sep=1.5pt]
\tikzstyle{originedge} = [draw = oldNodeEdge, thick, shorten <=0.5pt, shorten >=0.5pt]

\tikzstyle{targetnode-new} = [draw, circle, fill=newNodeEdge, inner sep=1.5pt]
\tikzstyle{targetnode-old} = [draw, circle, fill=oldNodeEdge, inner sep=1.5pt]
\tikzstyle{targetedge-new} = [draw = newNodeEdge, fill=newNodeEdge, thick, shorten <=0.5pt, shorten >=0.5pt]
\tikzstyle{targetedge-old} = [draw = oldNodeEdge, fill=oldNodeEdge, thick, shorten <=0.5pt, shorten >=0.5pt]
\tikzstyle{targetedge-new-nt} = [draw = newNodeEdge, fill=newNodeEdge, shorten <=0.3pt, shorten >=0.5pt]
\tikzstyle{targetedge-old-nt} = [draw = oldNodeEdge, fill=oldNodeEdge, shorten <=0.3pt, shorten >=0.5pt]

\tikzstyle{globalnode-new} = [draw, circle, fill=globalNode, inner sep=1.5pt]
\tikzstyle{globalnode-old} = [draw, circle, fill=globalNode, inner sep=1.5pt]
\tikzstyle{globaledge-new} = [draw = newNodeEdge, fill=newNodeEdge, thick, shorten <=1pt, shorten >=1pt]
\tikzstyle{globaledge-old} = [draw = oldNodeEdge, fill=oldNodeEdge, thick, shorten <=1pt, shorten >=1pt]

\tikzstyle{originnode-small} = [draw, circle, fill=black, inner sep=1pt]
\tikzstyle{originedge-small} = [draw, thick, shorten <=0.5pt, shorten >=0.5pt]

\tikzstyle{targetnode-old-small} = [draw, circle, fill=oldNodeEdge, inner sep=1pt]
\tikzstyle{targetnode-new-small} = [draw, circle, fill=newNodeEdge, inner sep=1pt]
\tikzstyle{targetedge-old-small} = [draw = oldNodeEdge, fill=oldNodeEdge, thick, shorten <=0.5pt, shorten >=0.5pt]
\tikzstyle{targetedge-new-small} = [draw = newNodeEdge, fill=newNodeEdge, thick, shorten <=0.5pt, shorten >=0.5pt]
\tikzstyle{globalnode-new-small} = [draw, circle, fill=newNodeEdge, inner sep=1pt]
\tikzstyle{globalnode-old-small} = [draw, circle, fill=oldNodeEdge, inner sep=1pt]
\tikzstyle{globaledge-new-small} = [draw = newNodeEdge, fill=newNodeEdge, thick, shorten <=0.5pt, shorten >=0.5pt]
\tikzstyle{globaledge-old-small} = [draw = oldNodeEdge, fill=oldNodeEdge, thick, shorten <=0.5pt, shorten >=0.5pt]

\tikzstyle{targetnode-old-big} = [draw, circle, fill=oldNodeEdge, inner sep=3pt]
\tikzstyle{targetnode-old-bigger} = [draw, circle, fill=oldNodeEdge, inner sep=2pt]
\tikzstyle{targetnode-new-bigger} = [draw, circle, fill=newNodeEdge, inner sep=2pt]
\tikzstyle{originnode-bigger} = [draw, circle, fill=black, inner sep=2.5pt]

\tikzstyle{vgnode} = [draw, circle, inner sep=1pt]
\tikzstyle{vgnode-big} = [draw, circle, inner sep=3pt]
\tikzstyle{vgedge} = [draw, shorten <=1pt, shorten >=1pt]

\tikzstyle{hcdnode} = [draw, circle, inner sep=1.5pt] \tikzstyle{hcdnode-big} = [draw, circle, inner sep=3pt]
\tikzstyle{hcdnode-bigger} = [draw, circle, inner sep=2pt]
\tikzstyle{hcdedge} = [draw, -stealth, shorten <=1pt, shorten >=1pt]
\tikzstyle{hcunode} = [draw, circle, fill=oldNodeEdge, inner sep=1.5pt] \tikzstyle{hcunode-big} = [draw, circle, fill=oldNodeEdge, inner sep=3pt] 
\tikzstyle{hcunode-bigger} = [draw, circle, fill=oldNodeEdge, inner sep=2pt] 
\tikzstyle{hcunode-narrow} = [draw, circle, fill=oldNodeEdge, inner sep=0.5pt] \tikzstyle{hcuedge} = [draw = oldNodeEdge, fill=oldNodeEdge, thick, shorten <=0.5pt, shorten >=0.5pt] 

\tikzstyle{crossedge} = [draw = newNodeEdge, fill=newNodeEdge, shorten <=0.5pt, shorten >=0.5pt] \tikzstyle{witnessedge} = [draw = orange!70, line width=3pt, shorten <=0.5pt, shorten >=0.5pt]

\usetikzlibrary{arrows,backgrounds,calc}

\pgfdeclarelayer{background}
\pgfsetlayers{background,main}

\tikzstyle{redEdge} = [draw = red!40, line width=5pt, shorten <=0.5pt, shorten >=0.5pt]
\tikzstyle{yellowEdge} = [draw = yellow!50, line width=5pt, shorten <=0.5pt, shorten >=0.5pt]
\tikzstyle{brownEdge} = [draw = blue!40, line width=5pt, shorten <=0.5pt, shorten >=0.5pt]

\definecolor{iltisBeige1}{HTML}{fef6ee}
\definecolor{iltisBeige2}{HTML}{e3d5c8}
\definecolor{iltisBeige3}{HTML}{cfbeb0}
\definecolor{iltisBeige4}{HTML}{bfac9b}

\definecolor{iltisGrey1}{HTML}{edebe8}
\definecolor{iltisGrey2}{HTML}{d9d6d2}
\definecolor{iltisGrey3}{HTML}{c2bfbc}
\definecolor{iltisGrey4}{HTML}{a6a4a1}

\definecolor{iltisLightGreen1}{HTML}{f4ffe9}
\definecolor{iltisLightGreen2}{HTML}{d4f7b2}
\definecolor{iltisLightGreen3}{HTML}{bde697}
\definecolor{iltisLightGreen4}{HTML}{a3d177}

\definecolor{iltisGreen1}{HTML}{cfffe1}
\definecolor{iltisGreen2}{HTML}{a2e8bd}
\definecolor{iltisGreen3}{HTML}{86d1a2}
\definecolor{iltisGreen4}{HTML}{6fbf8d}

\definecolor{iltisYellow1}{HTML}{fef2d0}
\definecolor{iltisYellow2}{HTML}{ffe3a2}
\definecolor{iltisYellow3}{HTML}{ffd77d}
\definecolor{iltisYellow4}{HTML}{f2c55e}

\definecolor{iltisRed1}{HTML}{ffe9e6}
\definecolor{iltisRed2}{HTML}{eda498}
\definecolor{iltisRed3}{HTML}{e08475}
\definecolor{iltisRed4}{HTML}{c26e60}

\definecolor{iltisOrange1}{HTML}{ffdfb3}
\definecolor{iltisOrange2}{HTML}{ffc97d}
\definecolor{iltisOrange3}{HTML}{ebb467}
\definecolor{iltisOrange4}{HTML}{e0a34c}

\definecolor{iltisCyan1}{HTML}{e0fffe}
\definecolor{iltisCyan2}{HTML}{b4e0df}
\definecolor{iltisCyan3}{HTML}{95c7c5}
\definecolor{iltisCyan4}{HTML}{81b3b0}

\definecolor{iltisBlue1}{HTML}{cce8ff}
\definecolor{iltisBlue2}{HTML}{8db8d9}
\definecolor{iltisBlue3}{HTML}{6f9abd}
\definecolor{iltisBlue4}{HTML}{5c86a8}
\definecolor{iltisBlue5}{HTML}{2e5b80}

\definecolor{iltisViolet1}{HTML}{ede6ff}
\definecolor{iltisViolet2}{HTML}{b1a5cc}
\definecolor{iltisViolet3}{HTML}{9b8eba}
\definecolor{iltisViolet4}{HTML}{8578a6}

\newcommand{\Clique}{\algorithmicProblem{Clique}}
\newcommand{\IS}{\algorithmicProblem{IndependentSet}}
\newcommand{\VC}{\algorithmicProblem{VertexCover}}
\newcommand{\FVS}{\algorithmicProblem{FeedbackVertexSet}}

\newcommand{\ThreeClique}{\algorithmicProblem{$3$-Clique}}
\newcommand{\FourClique}{\algorithmicProblem{$4$-Clique}}

\newcommand{\DS}{\algorithmicProblem{DominatingSet}}

\newcommand{\targetlabel}[2][]{\ifthenelse{\isempty{#1}}{\scriptsize\ensuremath{#2^\star}}{\scriptsize\ensuremath{(#2^\star\!\!,\!#1)}}}

\newcommand{\tupletargetlabel}[2][]{\ifthenelse{\isempty{#1}}{\scriptsize\ensuremath{(\!\{#2\},\!1)}}{\scriptsize\ensuremath{(#2,\!#1)}}}

\usepackage{tipa}
\usetikzlibrary{arrows}
\pgfdeclarelayer{backgroundA}
\pgfdeclarelayer{backgroundB}
\pgfdeclarelayer{background}
\pgfdeclarelayer{foreground}
\pgfsetlayers{background,backgroundB,backgroundA,main,foreground}

\newcommand{\borderradius}{1pt}
\newcommand{\borderwidth}{.8pt}

\definecolor{iltisheader}{rgb}{1,.89,.635}
\definecolor{iltistask}{rgb}{1,.95,.816}
\definecolor{iltisedge}{rgb}{.945,.725,.333}
\definecolor{iltispopupheader}{rgb}{.75,.675,.607}

\usetikzlibrary{decorations.text,decorations.footprints,decorations.shapes,shapes.symbols}

\usepackage{pgfplots}
\pgfplotsset{compat=1.16}

\usepgfplotslibrary{dateplot}
\usetikzlibrary{pgfplots.dateplot}

\usepackage[capitalise,noabbrev]{cleveref}

\begin{document}

\title{Tool-Assisted Learning of Computational Reductions}

\author{Tristan Kneisel}
\email{tristan.kneisel@rub.de}
\affiliation{\institution{Ruhr University Bochum}
  \streetaddress{Universitätsstraße 150}
  \city{Bochum}
  \country{Germany}
}
\author{Elias Radtke}
\email{elias.radtke@rub.de}
\affiliation{\institution{Ruhr University Bochum}
  \streetaddress{Universitätsstraße 150}
  \city{Bochum}
  \country{Germany}
}
\author{Marko Schmellenkamp}
\email{marko.schmellenkamp@rub.de}
\affiliation{\institution{Ruhr University Bochum}
  \streetaddress{Universitätsstraße 150}
  \city{Bochum}
  \country{Germany}
}
\author{Fabian Vehlken}
\email{fabian.vehlken@rub.de}
\affiliation{\institution{Ruhr University Bochum}
  \streetaddress{Universitätsstraße 150}
  \city{Bochum}
  \country{Germany}
}
\author{Thomas Zeume}
\email{thomas.zeume@rub.de}
\affiliation{\institution{Ruhr University Bochum}
  \streetaddress{Universitätsstraße 150}
  \city{Bochum}
  \country{Germany}
}

\begin{abstract}
    Computational reductions are an important and powerful concept in computer science. However, they are difficult  for many students to grasp. In this paper, we outline a concept for how the learning of reductions can be supported by educational support systems. We present an implementation of the concept within such a system, concrete web-based and interactive learning material for reductions, and report on our experiences using the material in a large introductory course on theoretical computer science.
\end{abstract}



\maketitle

\section{Introduction}\label{section:intro}

A \emph{computational reduction} is a computable function $f$ mapping source instances $a$ of an algorithmic problem $A$ to target instances $f(a)$ of an algorithmic problem $B$ with the property that

\begin{itemize}
 \item[] $a$ is a positive instance of $A$ \\ \hspace*{1cm}if and only if $f(a)$ is a positive instance of $B$
\end{itemize}

Intuitively, if the algorithmic problem $A$ reduces to $B$ via such a function, then it is ``at most as hard'' as $B$. Depending on the context, additional conditions are placed on reductions, such as polynomial-time computability.

Computational reductions are an important concept that recurs in several areas of computer science, such as in algorithms (as an abstraction of calls to sub-algorithms), in computability theory (for establishing undecidability of algorithmic problems), in complexity theory (for establishing NP-hardness of properties), and in SAT solving (as an abstraction for encoding procedures) \cite{GareyJ1979}. 

Because of their importance, computational reductions are often covered in mandatory or elective courses within undergraduate computer science curricula (see, for example, \cite{ACM2013,GI2016}). At our university, for example, computational reductions are part of a broad introductory course to theoretical computer science in the second year of a bachelor's degree, which covers material on formal languages, foundations of computability theory, and foundations of complexity theory.

In our experience, many students find computational reductions difficult and struggle to understand them. This often results in students being reluctant to even attempt exercises involving reductions because they are perceived as too difficult.

In this paper we address the question:
\begin{itemize}
 \item[] \textit{How can tools support the learning of computational reductions in large introductory courses?} 
\end{itemize}
Tool support ideally addresses both (1) students' difficulties in understanding and applying the concept of reductions and (2) students' reluctance due to the (perceived) difficulty of the subject. 

In classical (analogue) computational reduction assignments, several types of exercises are often used, including exercises for (A) understanding the algorithmic problems involved, (B) exploring existing reductions via examples, and (C) designing reductions between algorithmic problems. These exercises are often combined into multi-step exercises in which students first explore two concrete algorithmic problems and then reduce one to the other.

Existing tool support for learning reductions focuses on exercises of types (A) and (B) (see related work below). This is not surprising, as exercises and solution attempts of these types are easy to represent graphically, and it is straightforward to assess the correctness of student attempts and compute feedback. While exercises of type (C) have been addressed by some tools --- mainly by having students write reductions in a general-purpose programming language --- none of the tools focus on how to actually support students in learning to design reductions. There is also no tool support for multi-step exercises for learning computational reductions. 

\tikzstyle{mnode}=[
  circle,
  fill=iltisBlue3, 
  draw=black!80,
  minimum size=5pt, 
  inner sep=0pt,
]

\tikzstyle{uEdge}=[
thick, 
draw=black,
]

\tikzstyle{dEdge}=[
  -latex', thick, 
  shorten >=3pt, 
  shorten <=3pt,
  draw=black!80,
]

In particular, the lack of support for exercises  of type (C) is unfortunate, because such exercises are a challenge for most students. The likely reason for this is that the design of reductions does not usually follow a straightforward path, but requires some creativity on the part of the student. To teach such creatively demanding tasks, research in cognitive science suggests that it is helpful to teach novices how experts approach a problem, see \cite{HendersonMS2015} for a discussion and further references. 
When searching for a reduction, a typical approach for experts is to try sequentially a number of building blocks that they have previously encountered in the context of other reductions \cite{GareyJ1979}. 
An example of such a building block is provided by the standard reduction from the problem of finding a directed Hamiltonian cycle to finding an undirected Hamiltonian cycle. This reduction transforms a directed graph into an undirected graph by mapping each node \scalebox{0.7}{\hspace{-2mm}\begin{tikzpicture}[baseline=(1b)]
    \node[] (1a) at (-0.8,0.2){};
    \node[] (1b) at (-0.8,-0.2){};
    \node[] (3a) at (.8,0.2){};
    \node[] (3b) at (.8,0){};
    \node[] (3c) at (.8,-0.2){};
    \node[mnode, label=below:{\scriptsize $v$}] (2) at (0.0,0){};
    \draw[dEdge, very thick, shorten >=0pt, shorten <=0pt] (1a) -- (2);
    \draw[dEdge, very thick, shorten >=0pt, shorten <=0pt] (1b) -- (2);
    \draw[dEdge, very thick, shorten >=0pt, shorten <=0pt] (2) -- (3a);
    \draw[dEdge, very thick, shorten >=0pt, shorten <=0pt] (2) -- (3b);
    \draw[dEdge, very thick, shorten >=0pt, shorten <=0pt] (2) -- (3c);
    \end{tikzpicture}}\hspace{-1mm} to a small gadget \scalebox{0.8}{\hspace{-2mm}\begin{tikzpicture}[baseline=(tmp)]
    \node[] (tmp) at (-0.5,-.150){};
    \node[] (1a) at (-0.5,0.2){};
    \node[] (1b) at (-0.5,-0.2){};
    \node[] (3a) at (1.5,0.2){};
    \node[] (3b) at (1.5,0){};
    \node[] (3c) at (1.5,-0.2){};
    \node[mnode, label=below:{\scriptsize $v_\text{in}$}] (1) at (0,0){};
    \node[mnode, label=below:{\scriptsize $v$}] (2) at (0.5,0){};
    \node[mnode, label=below:{\scriptsize $v_\text{out}$}] (3) at (1,0){};
    \draw[uEdge, very thick, shorten >=0pt, shorten <=0pt] (1) -- (2);
    \draw[uEdge, very thick, shorten >=0pt, shorten <=0pt] (2) -- (3);
    \draw[uEdge, very thick, shorten >=0pt, shorten <=0pt] (1a) -- (1);
    \draw[uEdge, very thick, shorten >=0pt, shorten <=0pt] (1b) -- (1);
    \draw[uEdge, very thick, shorten >=0pt, shorten <=0pt] (3) -- (3a);
    \draw[uEdge, very thick, shorten >=0pt, shorten <=0pt] (3) -- (3b);
    \draw[uEdge, very thick, shorten >=0pt, shorten <=0pt] (3) -- (3c);
\end{tikzpicture}}\hspace{-1.1mm}. 
Such small graph gadgets --- such as replacing nodes or edges, or introducing a small global graph with some properties --- are typical building blocks when designing reductions. We envision helping students to learn how to design reductions by exposing them to such gadgets in given reductions, and then having them design similar reductions themselves.\footnote{Gal-Ezer and Trakhtenbrot take a superficially related approach by trying to identify typical patterns in reductions in undecidability proofs \cite{Gal-EzerT16b}.} Of course, in this approach, students do not directly see how to come up with new building blocks --- a task required in a research environment. But they do learn how to construct reductions in a simplified framework.

\subsubsection*{Contributions} We implement tool support for exercises of types (A) -- (C) within the educational support system \anonymize{\Iltis}{\anon} \anonymize{\cite{GeckLPSVZ18,GeckQSSTVZ21,SchmellenkampVZ24}}{[\anonshort]}, design interactive teaching material that supports the learning of computational reductions with multi-step exercises, and report on experiences of using the material in a large course in which $\sim 175$ students actively engaged with the material.

To support exercises of types (A) -- (C) in \anonymize{\Iltis}{\anon}, we design and implement two new reduction-specific interactive task types:
\begin{itemize}
    \item Exercises of type (A) and (B) require students to be able to construct graphs and colour their nodes and edges. The first type of interactive task is a \emph{general graph construction task} which allows students to construct (parts of) graphs and colour sets of nodes or edges of graphs subject to an expressive set of constraints. 

    \item For exercises of type (C), we follow the approach outlined above. The second type of interactive task is a \emph{reduction design task}, which allows students to construct graph-based computational reductions with variation in the type of building blocks that students can use. This task implements parts of the recently introduced theoretical \emph{Cookbook reduction framework}, which provides a specification language for reductions using building blocks such as those outlined above, and has nice algorithmic properties~\anonymize{\cite{GrangeVVZ2024}}{[\anonshort]\footnote{For anonymity reasons, we omit the reference.}}.

\end{itemize}

\subsubsection*{Related work}

Tools for supporting the learning of computational reductions mostly cover exercises of type (A) and (B) from above. 
The most sophisticated framework for illustrating reductions is \emph{GraphBench}, which allows visualising the construction of reduced instances for several reductions, exploring reductions by changing source instances and observing the effect on targets, and many other features~\cite{Brandle06}. Other visualisations have been developed \cite{Maji15,Vegdahl2014,Crescenzi10,Pape98}. The understanding of formal problems is supported by the system in \cite{Maji15}, which also supports the translation of solution candidates for problems according to a given reduction.  An approach to exercises of type (C) inspired by automated programming exercises is taken by Engström and Kann~\mbox{\cite{EnstromK17,CrescenziEK13}}. Zhang et al.~\cite{ZhangHD22} extend this approach by introducing a language for specifying reductions and testing them on random problem instances.

To the best of our knowledge, there are no tools that support students in learning \emph{how} to construct reductions. One of the reasons is that this requires a deep theoretical understanding, since it is generally undecidable to decide whether a computational reduction provided by a student is indeed a reduction from an algorithmic problem $A$ to a problem $B$. A specification language that allows students to construct reductions thus has to tread a fine line between being sufficiently expressive and being algorithmically accessible. The Cookbook reduction framework of \anonymize{Grange et al.}{\anon} provides such a language \anonymize{\cite{GrangeVVZ2024}}{[\anonshort]}.

\subsubsection*{Outline of the paper}
In \cref{section:workflows}, we outline which types of (multi-step) exercises should be supported by a tool for learning reductions. Then, in \cref{section:tasks}, we describe our implementation of these ideas within the educational support system \anonymize{\Iltis}{\anon}. In \cref{section:classroom} we report on our experiences and conclude in \cref{section:conclusion}.

\section{Exercises and Multi-step Exercises}\label{section:workflows}
In this section we analyse the requirements for tool support for computational reductions; in the next section we report on our implementation of these requirements within the educational support system \Iltis. We aim to support students with exercises for (A) understanding the algorithmic problems involved, (B) exploring existing reductions via examples, and (C) designing reductions between computational problems. We discuss these types of exercises and how they can be flexibly combined into multi-step exercises.

\subsubsection*{Fine-granular exercise types}
Several fine-grained types of exercises can be used to train each of (A) -- (C):

\begin{enumerate}[label=(\Alph*)]
 \item Understanding  algorithmic problems
        \begin{enumerate}[label=(\arabic*)]
            \item Indicating a solution for a positive instance
            \item Indicating why a solution candidate is not a solution
            \item Constructing an instance which satisfies certain constraints (e.g. of a certain size, with additional properties, etc.)
        \end{enumerate}
 \item Understanding reduction candidates
        \begin{enumerate}[label=(\arabic*)]
            \item Applying a reduction to an input instance
            \item Providing a counterexample for a reduction candidate
            \item Transferring a solution candidate according to a reduction
        \end{enumerate}\item Designing computational reductions
        \begin{enumerate}[label=(\arabic*)]
            \item Designing a reduction using a global gadget
            \item Designing a reduction using an edge gadget
            \item Designing a reduction using a node gadget
\end{enumerate}
\end{enumerate}

The following examples illustrate some aspects of these fine-grained exercises.
\newcommand{\understandingproblemslabel}{(A)\xspace}
\newcommand{\understandingreductionslabel}{(B)\xspace}
\newcommand{\understandingnonreductionslabel}{(C)\xspace}
\newcommand{\constructingreductionslabel}{(D)\xspace}

\begin{example}[Understanding the problem \VC]\label{example:understand-vc}
The algorithmic problem \VC asks, given an undirected graph $ G $ and a number $ k $, whether there exist $ k $ nodes in $ G $ that are adjacent to all edges. Such a set of nodes is called a \emph{vertex cover} of~$G$. Understanding the \VC problem boils down to understanding when a set of nodes is a vertex cover. Some possible instantiations of exercises  (A) for understanding \VC are the following:
\begin{itemize}
    \item[(A1)] ``Select three nodes that form a vertex cover of the following graph.''
    \item[(A2)] ``Select an edge that shows that the nodes highlighted in the following graph do not form a vertex cover.''
    \item[(A3a)] ``Construct a graph with 7 edges that has a vertex cover of size three.''
    \item[(A3b)] ``Construct a graph that has a vertex cover of size two but no dominating set of size two.''
\end{itemize}
Here (A1) and (A2) are used to explore positive and negative instances. Exercises (A3a) and (A3b) also explore instances, but with different constraints. Exercises of the last type can be used to explore differences between algorithmic problems. 

These exercises require students to view and construct graphs, to highlight and select nodes and edges, and to check flexible constraints on graphs and selections.\qed
\end{example}

\begin{example}[Understanding mappings from \VC to \DS]
 The algorithmic problem \DS asks, given an undirected graph $ G $ and a number $ k $,  whether there exists a dominating set of size at most $ k $, i.e. a set $U$ of nodes such that all nodes $v$ of $G$ are in $U$ or adjacent to a node in $U$. Instantiations of the exercises of type (B) can be used to explore candidates for reductions from \VC:
 \begin{itemize}
  \item[(B1)] ``Apply the candidates $f_1$ and $f_2$ for reductions from \VC to \DS to the following graph $G$.''
  \item[(B2)] ``Show that $f_1$ is not a reduction from \VC to \DS by giving a counterexample.''
  \item[(B3)] ``As we have seen, $f_2$ is a reduction from \VC to \DS, since solutions of \VC are mapped to solutions of \DS, and vice versa. Select a solution for the \DS instance $f_2(G)$ which is induced by the following solution for the \VC instance G.''
 \end{itemize}
 Such exercises can be interspersed with multiple choice questions asking students whether they think a reduction candidate is correct or not, and in either case how they would prove it. Additional information like proof ideas is provided through text passages.
 
 The exercises for viewing, constructing, highlighting and selecting (parts of) graphs have similar requirements as stated in \cref{example:understand-vc}.
\qed 
\end{example}

\begin{example}[Designing a reduction from \VC to \FVS] \label{example:reduction-vc-fvs}
    The algorithmic problem \FVS asks, given an undirected graph $G$ and a natural number $k$, whether it is possible to obtain a graph without cycles by removing at most $k$ nodes from $G$. The standard reduction from  \VC to \FVS replaces edges \scalebox{0.85}{\tikz[baseline={([yshift=-0mm]current bounding box.center)}]{
 \node[mnode, label={[shift={(0,0)}, label distance=-2pt]below:{\scriptsize  $ u $}}] (v2) at (-0.25,0.0) {};
 \node[mnode, label={[shift={(0,0)}, label distance=-2pt]below:{\scriptsize  $ v $}}] (v1) at (0.25,0.0) {};
 \draw[uEdge] (v1) edge (v2);
 }}
of $G$ with a small triangle gadget graph \scalebox{0.85}{\tikz[baseline={([yshift=-0mm]current bounding box.center)}]{
\node[mnode, label={[shift={(0,0)}, label distance=-2pt]below:{\scriptsize  $u$}}] (v2) at (-0.25,0) {};
 \node[mnode, label={[shift={(0,0)}, label distance=-2pt]below:{\scriptsize $v$}}] (v1) at (0.25,0) {};
 \node[mnode, label={[shift={(0,0)}, label distance=-2pt]right:{\scriptsize  $w_{uv}$}}] (v12) at (0,0.25) {};
 \draw[uEdge] (v1) edge (v2) edge (v12) 
 	(v2) edge (v12);
}}\hspace{-1mm}.    
This is an example of an edge gadget;  another  typical building block for reductions.

A possible instantiation of an exercise of type (C3) is as follows:
\begin{itemize}
 \item[(C3)] ``Design an edge gadget that reduces \VC to \FVS.'' 
\end{itemize}
Such exercises require a nice way for students to input reductions, as well as algorithms for checking the correctness of reductions and providing feedback. We emphasize that the triangle gadget from above is not the only gadget that can be used to reduce \VC to \FVS. Therefore, for checking correctness and providing feedback, the gadgets provided by students need to be carefully analysed. 
\qed
\end{example}

\subsubsection*{Multi-step Exercises} 
To guide students, it is essential that exercises can be flexibly combined into multi-step exercises. For example, a multi-step exercise that introduces students to edge gadget reductions might look as follows:
\begin{itemize}
 \item[(1)] Exploring an edge gadget reduction from \VC to \DS
	\begin{itemize}
	 \item[(a)] Understanding the problem \VC
	 \item[(b)] Understanding the problem \DS
	 \item[(c)] Understanding an edge gadget reduction $f$ from \VC to \DS
	\end{itemize}
 \item[(2)] Designing an edge gadget reduction from \VC to \FVS
	\begin{itemize}
	 \item[(a)] Understanding the problem \FVS
	 \item[(b)] Designing an edge gadget reduction $g$ from \VC to \FVS
	 \item[(c)] Applying $g$ to an input instance
	\end{itemize}
\end{itemize}

One requirement of such multi-step exercises can be seen in (2b) and (2c), where a reduction $g$ designed by students as part of (2b) is used in the subsequent exercise (2c).

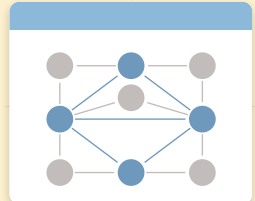
\begin{figure*}
    \tikzset{
        task/.style={
            rectangle split,
            rectangle split parts=3,
            rectangle split part fill={iltisblue1, iltisblue0, iltisblue1},
            rectangle split part align={left, center,left},
            rectangle split draw splits=false,
            every one node part/.style={font=\scriptsize},
            every two node part/.style={font=\footnotesize},
            every three node part/.style={font=\scriptsize},
            rounded corners=3pt,
            inner xsep=5pt,
            inner ysep=4pt,
            align=center,
            line width=.3pt,
        },
        taskedge/.style={
            ->,
            draw,
            line width = 2pt,
            iltisblue2,
            shorten <=2pt,
            shorten >=2pt
        },
        exercise/.style={
            node distance = 0.3cm,
            every node/.style={task},
            every edge/.append style={taskedge},
        },
        connector/.style={
            decoration={footprints,foot of=felis silvestris,foot length=5pt,stride length=10pt,foot sep=1pt},
            decorate,
            iltisedge,
        },
        comment/.style={
            draw=iltisheader,
            fill=iltistask,
            rectangle,
            rounded corners=\borderradius,
            line width=\borderwidth,
            inner xsep=5pt,
            inner ysep=4pt,
            node font=\footnotesize,
        }
    }

\newcommand{\includeScreenshot}[6]{
        \includegraphics
        [width=#2,clip,trim=#3 #4 #5 #6,]{#1}}

\newcommand{\includeIltisScreenshotWithBorderInTikz}[8]{
        \node (#1)[screenshot,#2] {\includeScreenshot{#3}{#4}{#5}{#6}{#7}{#8}};
        \node [white screenshot border] at (#1) {\phantom{\includeScreenshot{#3}{#4}{#5}{#6}{#7}{#8}}};
        \node [screenshot border] at (#1) {\phantom{\includeScreenshot{#3}{#4}{#5}{#6}{#7}{#8}}};
    }
\newcommand{\includeIltisScreenshotWithBorderInTikzDefaultTrim}[5]{\includeIltisScreenshotWithBorderInTikz{#1}{#2}{#3}{#4}{13mm}{#5}{13mm}{52mm}}
\newcommand{\includeIltisScreenshotWithBorderInTikzAlreadyCropped}[4]{\includeIltisScreenshotWithBorderInTikz{#1}{#2}{#3}{#4}{0mm}{0mm}{0mm}{0mm}}

    \tikzset{screenshot/.style={
        inner sep=0,outer sep=0,
        align=center, }}

    \tikzset{screenshot border/.style={
        draw=iltisheader,
        rounded corners=\borderradius,
        line width=\borderwidth,
        inner sep=0,outer sep=0,
        inner xsep=-1.5pt,
        align=center, }}

    \tikzset{white screenshot border/.style={
        draw=white,
        line width=\borderwidth,
        inner sep=0,outer sep=0,
        inner xsep=-1.5pt,
        align=center, }}

    \newlength{\fullwidth}
    \newlength{\taskwidth}
    \newlength{\sswidth}
    \newlength{\commentwidth}
    \newcommand*{\goalsep}{1mm}
    \newcommand*{\tasksep}{5mm}
    \newcommand*{\boxsep}{9mm}
    \newcommand*{\sssep}{2mm}
    \newcommand*{\colsep}{8mm}
    \setlength{\fullwidth}{178.6mm} \addtolength{\fullwidth}{-2mm}
    \setlength{\taskwidth}{.5\fullwidth}
    \addtolength{\taskwidth}{-4mm} \setlength{\sswidth}{\taskwidth}
    \addtolength{\sswidth}{2mm} \addtolength{\sswidth}{-4mm} \setlength{\commentwidth}{\sswidth}
    \addtolength{\commentwidth}{-3.5mm}
\scalebox{1}{
    \begin{tikzpicture}[
            screenshot/.append style={node distance=1mm and 1mm},
            comment/.append style={node distance=1mm and 1mm},
            goal-description/.style={
                minimum width=\taskwidth,
font=\footnotesize,
                inner xsep=0mm,
                },
            task-description/.style={
                minimum width=\taskwidth,
draw=iltisblue2,
                fill=iltisblue0,
                font=\footnotesize,
                rounded corners=3pt,
                },
            goal-box/.style={
                minimum width=\sswidth,
                fill=iltisblue1,
                font=\footnotesize,
                rounded corners=3pt,
            },
            task-descr-edge/.style={
                ->,
                draw,
                line width = 2pt,
                iltisblue2,
                shorten <=2pt,
                shorten >=2pt,
            },
            box-edge/.style={
                task-descr-edge,
            },
        ]

\node (assignment) [goal-description, minimum width=\fullwidth]
            {\textbf{Assignment}: First explore an edge gadget reduction from \textsc{VertexCover} to \textsc{DominatingSet}, then design a similar reduction from \textsc{VertexCover} to \textsc{FeedbackVertexSet}.};

\node (explore-f1) [goal-description,below=4mm of assignment.south west, anchor=north west]
            {\textbf{(1) Exploring an edge gadget reduction} \scalebox{.9}{from \textsc{VertexCover} to \textsc{DominatingSet}}};
        \node (explore-f1-vc) [task-description,below=\goalsep of explore-f1] 
            {\begin{minipage}{7.5cm}
                (a) Understanding the problem \textsc{VertexCover}
                \begin{itemize}[leftmargin=6mm]
                    \item[$\triangleright$] Select a vertex cover
                    \item[$\triangleright$] Show that nodes do not form a vertex cover 
                \end{itemize}
            \end{minipage}};
        \node (explore-f1-ds) [task-description,below=\tasksep of explore-f1-vc] 
            {\begin{minipage}{7.5cm}
								(b) Understanding the problem \textsc{DominatingSet}              
             \end{minipage}
};
        \node (explore-f1-reduction) [task-description,below=\tasksep of explore-f1-ds] 
            {\begin{minipage}{7.5cm}
                (c) Understanding a reduction $f$ from \textsc{VertexCover} to \textsc{DominatingSet}
                \begin{itemize}[leftmargin=6mm]
                    \item[$\triangleright$] Apply $f$ to a \textsc{VertexCover} instance
                    \item[$\triangleright$] Transfer a solution candidate from \textsc{VertexCover} to  \textsc{DominatingSet}
                \end{itemize}
            \end{minipage}};

\node (design-f) [goal-description,right=\colsep of explore-f1.north east, anchor=north west] 
            {\textbf{(2) Designing an edge gadget reduction} \scalebox{.9}{from \textsc{VertexCover} to \textsc{FeedbackVertexSet}}};
        \node (design-f-fvs) [task-description,below=\goalsep of design-f] 
            {\begin{minipage}{7.5cm}
                (a) Understanding the problem \textsc{FeedbackVertexSet}
                \begin{itemize}[leftmargin=6mm]
                    \item[$\triangleright$] Select a feedback vertex set
                    \item[$\triangleright$] Show that nodes do not form a feedback vertex set
                    \item[$\triangleright$] Construct a graph such that all feedback vertex sets have size at least 3
                    \item[$\triangleright$] \dots
                \end{itemize}
            \end{minipage}};
        \node (design-f-construct) [task-description,below=\tasksep of design-f-fvs] 
            {\begin{minipage}{7.5cm}
								(b) Design a reduction \(g\) from \textsc{VertexCover} to \textsc{FeedbackVertexSet}
             \end{minipage}
						};
        \node (design-f-apply) [task-description,below=\tasksep of design-f-construct] 
            {\begin{minipage}{7.5cm}
            (c) Apply \(g\) to a \VC instance
             \end{minipage}
};  

        \draw (explore-f1-vc) edge [task-descr-edge] (explore-f1-ds);
        \draw (explore-f1-ds) edge [task-descr-edge] (explore-f1-reduction);
        \draw (design-f-fvs) edge [task-descr-edge] (design-f-construct);
        \draw (design-f-construct) edge [task-descr-edge] (design-f-apply);

        \begin{pgfonlayer}{backgroundA}
            \node (assignment-box) [goal-box,fit={(assignment)}] {};
            \node (f1-box) [goal-box,fit={(explore-f1) (explore-f1-reduction)}] {};
            \node (f-box) [goal-box,fit={(design-f) (design-f-apply)}] {};
            \draw (f1-box) edge [box-edge] (f-box);
        \end{pgfonlayer}

        \includeIltisScreenshotWithBorderInTikz{ss-vc}{below=\sssep of f1-box}{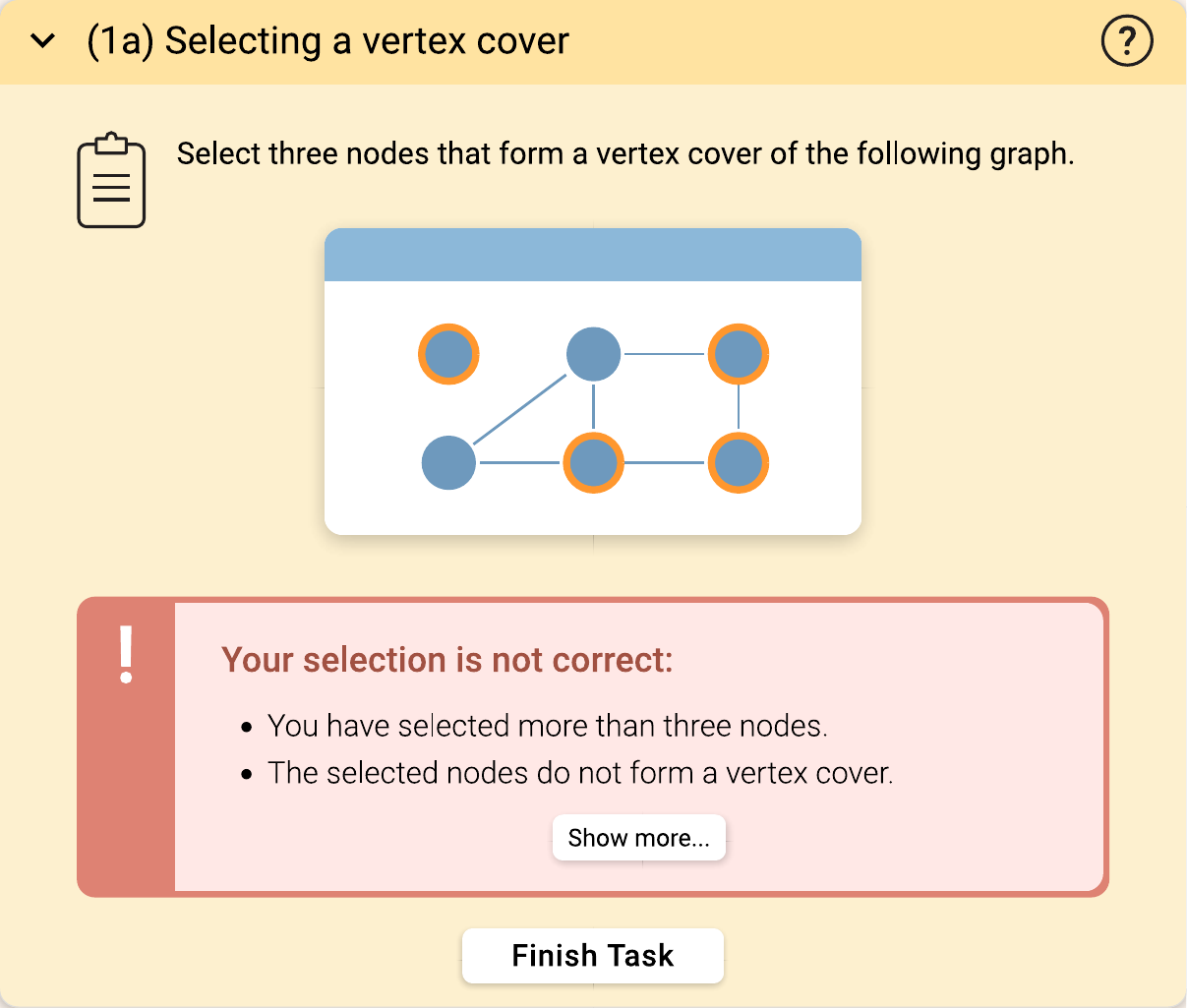}{\sswidth}{1mm}{1mm}{1mm}{1mm}
        \includeIltisScreenshotWithBorderInTikz{ss-transfer}{below=\sssep of ss-vc,yshift=-2mm}{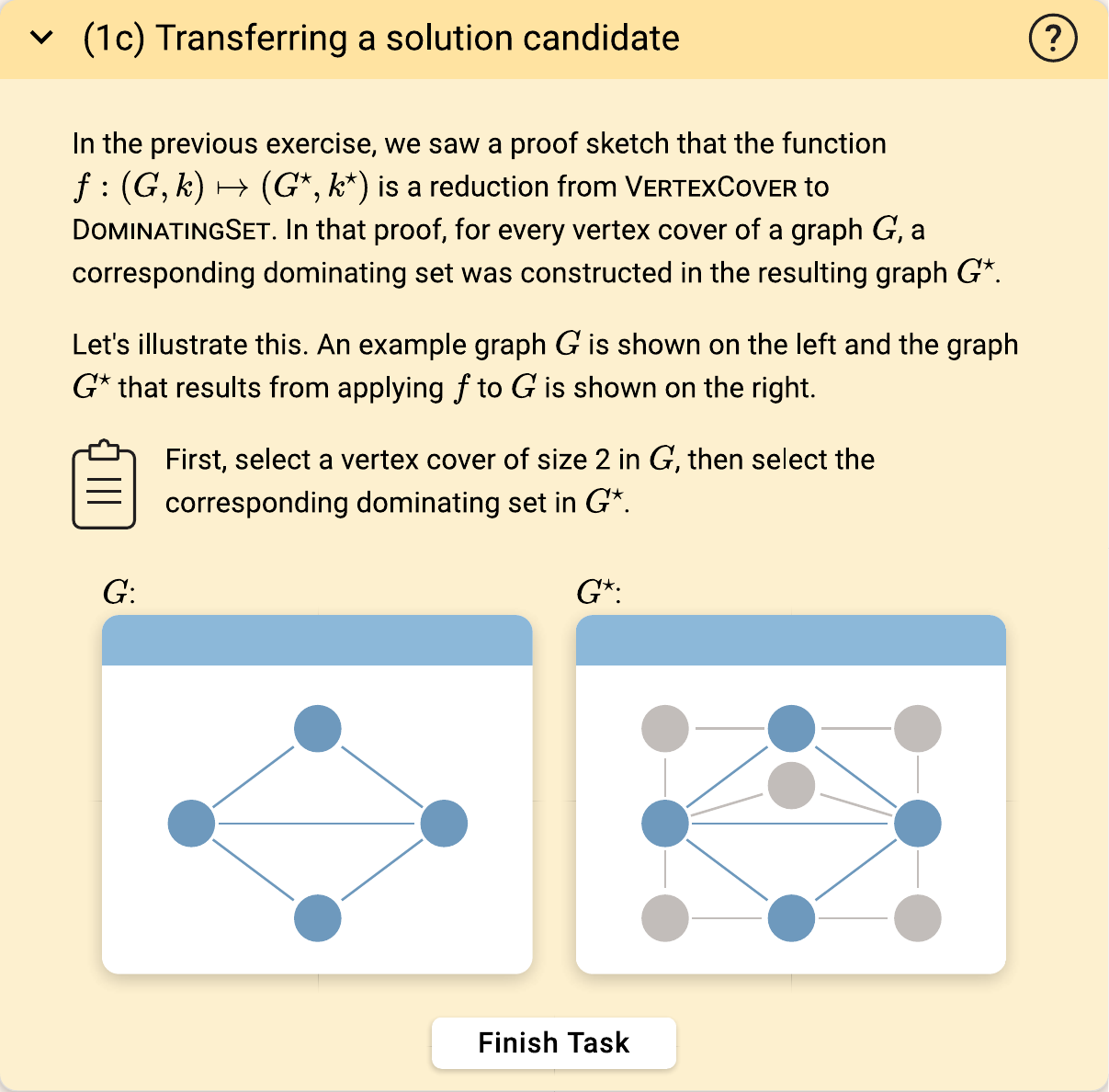}{\sswidth}{1mm}{1mm}{1mm}{1mm}
        \includeIltisScreenshotWithBorderInTikz{ss-reduction}{below=\sssep of f-box}{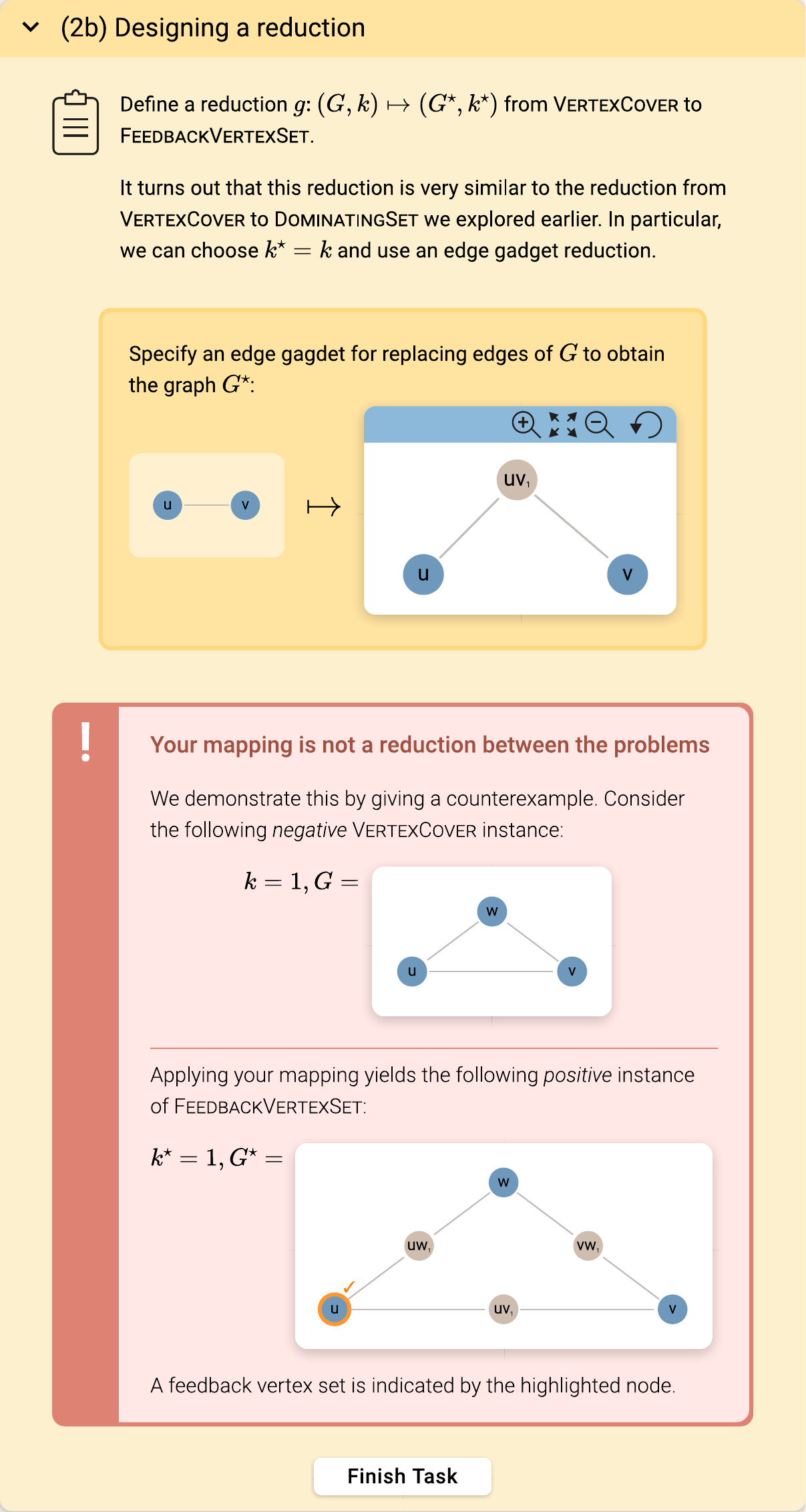}{\sswidth}{1mm}{1mm}{1mm}{1mm}

    \end{tikzpicture}
    }

\caption{Illustration of a multi-step exercise for constructing computational reductions between two problems. Some steps are illustrated by screenshots from the system. Parts of this illustration are adopted from \anonymize{\cite{SchmellenkampVZ24}}{[\anonshort]}.}\label{figure:reductions}
\end{figure*}

\section{Computational Reductions in \Iltis}
\label{section:tasks}

In the previous section, we saw that for supporting students in their learning of computational reductions, educational support systems require (R1) support for multi-step exercises; (R2) a graph construction task; and (R3) a reduction design task. We implement these requirements within the educational support system \Iltis.

The educational support system \Iltis allows teachers to flexibly build multi-step exercises from a portfolio of small, easily composable educational tasks. Each task is configurable by inputs -- either given explicitly or as the output of previous tasks -- and provides objects created by students within that task as outputs. The outputs can then be used by subsequent tasks. 

Thus, (R1) is a built-in feature of \Iltis, and we focus on (R2) and (R3) in the following.

\subsubsection*{A Graph Construction Task} 
The user interface of our graph construction task has a configurable graph widget which allows to view and construct graphs, as well as select/highlight nodes and edges within graphs. The widget can optionally be initialized with a teacher-specified graph. The task is illustrated in \cref{figure:reductions}.

In the backend, the task has a \emph{constraint framework} that allows teachers to specify complex constraints that constructed graphs, selections, and highlightings must satisfy. Several types of constraints are supported:
\begin{itemize}
 \item \emph{Isomorphy constraints:} The constructed graph is isomorphic to a specified graph.
 \item \emph{Cardinality constraints:} The number of (highlighted/selected) nodes or edges of a  graph satisfies some cardinality condition. An example of such a constraint is that ``between 2 and 4 nodes have been selected''.
 \item \emph{Logical constraints:} The graph and its selected/highlighted nodes  and edges satisfy a specified formula in first-order logic. An example of such a constraint is that a vertex cover has been selected, i.e. $\forall u \forall v [E(u,v) \to (S(u) \lor S(v))] $ where $E$ denotes the edge relation and $S$ denotes a unary relation containing the selected nodes. 
\end{itemize}
Constraints can be negated and combined by requiring that either all constraints, or at least one constraint, or none of a list of constraints should be satisfied. Generic feedback is generated that, if so, addresses which constraints are not satisfied.

\subsubsection*{A Reduction Design Task} 
There are two main challenges in developing a reduction design task. First, computational reductions are a difficult concept for many students. Therefore, it is essential that reductions can be specified in a simple and intuitive way in order to keep the external cognitive load for students low. Second, verifying the correctness of computational reductions in general is undecidable, even for severely restricted classes of reductions such as polynomial-time or even first-order definable reductions \anonymize{\cite{GrangeVVZ2024}}{[\anonshort]}.

We address both challenges by focusing on the following types of reductions:
\begin{itemize}
	\item \emph{Global gadget reductions:} Construct a target instance by (i) adding one graph gadget to the source instance, and (ii) connecting it appropriately. An example is the standard reduction from the problem of finding a clique of size 3 (\ThreeClique) to finding a clique of size 4 (\FourClique), which adds a single node to the source instance and connects it to all other nodes. 
	\item \emph{Edge gadget reductions:} Construct a target instance by replacing each (non-)edge in the source instance with a graph gadget. An example is the standard reduction from \VC to \FVS (see \cref{example:reduction-vc-fvs}).
	
\item \emph{Node gadget reductions:} Construct a target instance by replacing each node in the source instance with a graph gadget. An example is the standard reduction from the problem of finding a directed Hamiltonian cycle to finding an undirected Hamiltonian cycle (see the example in \cref{section:intro}).
\end{itemize}

For all three types of reductions, our reduction design task provides a user interface for graphically specifying gadgets and how they are integrated into the target instance. Optionally, to help students understand their reduction, the task also supports providing a sample source instance to which the specified reduction is applied. We refer to \cref{figure:reductions} for an illustration of this task. 

Unfortunately, even for these special types of reductions, automatic verification of attempts is a provably algorithmically hard task. For this reason, theoretical efforts are required to support a new reduction. Each of our supported reductions relies on a theoretical characterization (most of them from \anonymize{\cite{GrangeVVZ2024}}{[\anonshort]}) of when a gadget is correct and, if not, why it is not.

\begin{example}[Correct edge gadgets for reductions from \VC to \FVS]
We illustrate this idea with the characterization of correct edge gadgets for reducing \VC to \FVS.
An edge gadget replaces each edge $ (u, v) $ of the source instance by a gadget graph $G_{u,v}$ with distinguished nodes $u$ and $v$. Such an edge gadget for reducing \VC to \FVS (with the same $k$) is correct if and only if it satisfies the following two properties:
\begin{enumerate}
    \item $ G_{u,v} $ contains a cycle, and
    \item removing $ u $ or $ v $ from $ G_{u,v} $ results in a graph without cycles.
\end{enumerate}Intuitively speaking, property (1) is required, because if $ G_{u,v} $ does not contain a cycle, then any negative source instance $ (G, k) $ with fewer than $ k $ cycles is a counterexample.  When property (2) is violated it is easy to find positive source instances with negative target instances. These intuitions are used to provide feedback. \qed
\end{example}

We have used theoretical characterizations to implement, among others, global gadget reductions from \ThreeClique to \FourClique, node gadget reductions from directed to undirected Hamiltonian cycle, and edge gadget reductions from \VC to \FVS or \DS, and from \Clique to \IS. Feedback for solution attempts is provided by checking the properties of the corresponding characterization and, if not satisfied, a simple counterexample is derived.

\section{Use in the classroom}\label{section:classroom}

We used tool-based exercises for computational reductions described above in an introductory course on theoretical computer science. In the following, we describe the setting of the course (Section \ref{section:setting}), how reduction-related multi-step exercises have been used in the course (Section \ref{section:materialinclass}), and how the usage of the material was perceived by the students (Section \ref{section:evaluation}). We emphasize that a sound study of the effectiveness of our material is out of scope for this paper, instead we report on data that hints at its usefulness.

\subsection{Setting}\label{section:setting}

The course \emph{Foundations of Theoretical Computer Science} at \anonymize{Ruhr University Bochum, Germany}{\anon} is an introductory course in formal languages, automata theory, computability and complexity theory in the third semester of a bachelor's degree in computer science and two related degree programmes. The course spans 12 weeks and is divided equally into four parts: The first two parts focus on regular and context-free languages, the third on decidability and computability, and the fourth on complexity theory, in particular on P and NP and the relevance of NP-completeness. Reductions are introduced as a formal foundation for (un)decidability in part three and specialized to polynomial reductions as a basis for NP-hardness results in part four.

We shortly describe the organization of this course in the winter terms 2022/23 and 2023/24: (i) The content of the course is provided in lectures with slides, (ii) for practising the content, weekly homework assignment sheets are provided. Assignments are to be submitted partly in analogue form in groups of up to three students (paper-based assignments) and partly individually via the educational support system \anonymize{Iltis}{\anon} (web-based assignments). For points achieved on the assignments, a (small) bonus on the grade in the final exam is provided. Students can also use (iii) further assignments in the support system for free practice (practice assignments).

Per week, (I)  there are  two 90\,min lectures in which the lecture slides are presented, also (II) there are tutorial sessions in which solutions for the homework assignments (and also other assignments) are discussed. Additionally, (III) in the second half of the course, a help desk for asking questions is offered.

\subsection{Material}\label{section:materialinclass}

In winter term 2023/24, in addition to paper-based assignments, web-based assignments for computational reductions as described in this paper are  used for the first time, at three stages: 
\begin{description}
    \item[Assignment Workflow 1:] In Week 9, computational reductions are introduced in the context of computability. In the lecture, the first examples are for familiar graph problems, later reductions for problems involving Turing machines are introduced. In the web-based assignments, we focus on graph-based reductions. Students explore a reduction via a global gadget; they prove that a given (very similar) function is not a reduction; and they construct a global gadget reduction that is very similar to the one they explored, but uses different algorithmic problems. For each algorithmic problem, tasks for understanding the problem are offered. 

	\item[Assignment Workflow 2:] In Week 11, polynomial reductions are introduced. In the web-based assignments, we replicate Assignment Workflow 1, but use reductions with an edge gadget. \cref{figure:reductions} shows an excerpt of these assignments.

	\item[Assignment Workflow 3:] In a recap at the end of the course, we iterate the assignment workflows for non-edge gadget constructions. Here, designing a reduction is embedded into a multi-step exercise for proving NP-hardness of an algorithmic problem. This includes exercises for students to select the correct direction of the reduction and choosing the correct time complexity of the constructed reduction. All the algorithmic problems have been encountered before, so no exercises for understanding these problems are necessary.
\end{description}
The material for the workflows is provided under the link \anonymize{\url{https://iltis.cs.tu-dortmund.de/computational-reductions/}}{\anon}.

\subsection{Experience Report}\label{section:evaluation}

We first report on challenges when teaching reductions and then discuss our experiences on the usefulness of our system.

\paragraph{Challenges for teaching reductions}

Instructors who taught the course report that reductions are one of the hardest topics of the course. Even though several weeks are spent on reductions, many students struggle and give up on this topic. Interestingly, this is not limited to the task of designing reductions --- which is difficult and sometimes frustrating as it depends on experience and oftentimes a certain intuition about the problems involved --- but also effects tasks with significantly lower difficulty, e.\,g. applying a reduction to a given input --- that is, applying a function to a given input.
This can also be witnessed in exams, where sub-assignments that ask to provide positive/negative instances for problems and to apply functions to inputs are attempted less frequently than expected.

\paragraph{Experiences and Observations}
We discuss our experiences and observations when using the Assignment Workflows 1 -- 3.

\paragraph{(a) Providing a sense of achievement}

The material was used extensively. As a point of reference, the web-based assignments in multiple choice format, that were posed in the weeks surrounding the weeks with the reduction assignments, were completed by an average of 151 students (176 in Week 8, 162 in Week 10, 119 in Week 12).\footnote{The course was attended by $\sim 300$ students. At least one web-based assignment was completed by 311 students; 317 enrolled for the exam. Since assignments are not mandatory, the number of active students drops significantly over the semester.}
For the reduction-related assignments, the single steps of the Assignment Workflows~1 and~2 in Weeks 9 and 11 were completed by between 100 and 140 students (as Workflow~3 was not part of the bonus point scheme, it was attempted much less often).

From the students that completed the multi-step exercise for understanding the relevant problems, 88\,\% (Workflow 1) to 92\,\% (Workflow 2) also completed the assignments for designing a new reduction. These rates are considerably higher than rates estimated by teachers who grade paper-based assignments on reductions. 
This suggests that our tool provides students with a sense of achievement.
In addition, students only needed an average of 2.0 (Workflow 1) to 2.2 (Workflow 2) attempts on the reduction design task.  

From our data we cannot conclude whether this effect manifested in more students than usual attempting the paper-based assignments on reductions, i.\,e. whether our material had an overall motivating effect. We leave this for future work.

\paragraph{(b) Usefulness for students}

During the submission phase, after they designed a reduction, we asked students whether (i) they felt prepared to provide the reduction by the previous steps of the assignment workflows; (ii) they felt it was easy to input the reduction in the required format; and whether (iii) they felt prepared to write the reduction on paper after providing it in our task. For all questions, we used a 
semantic differential 
scale from 1 to 5 with 1 meaning \enquote{very easy}/\enquote{very well prepared} and 5 meaning \enquote{very hard}/\enquote{very badly prepared}. As only 8 students answered the survey for Workflow~3 in the recap section, we focus on Workflows 1 and 2 for which we have > 90 answers each.

First, we report on the answers to the survey after Workflow 1: For question (i), the average is 2.7, which means that on average, students felt only slightly prepared for the reduction by the previous workflow steps.
For question (ii), the average is 3.0, meaning that students found it moderately difficult to enter the reduction in the required format. 
Finally, for question (iii), the average is 3.3, meaning that the students did not really feel prepared to write the given reduction on paper in the usual notation.
Even though the students were more familiar with the type of reduction exercises in the second workflow, the average of the answers after Workflows~1 and~2 was very close for all questions (differences of up to 0.3).
For all questions we observed a high variability of the answers (the standard deviation was between 1.0 and 1.3).

As designing reductions is inherently difficult, it is not surprising that students tend to feel that --- even with tool support --- coming up with reductions is difficult. Yet, these ratings can be taken as an indication to improve the wording of the exercises and to provide more information on how to transfer reductions to paper.

\section{Conclusion and Outlook}
\label{section:conclusion}
We outlined how support for learning computational reductions can be integrated into educational support systems such as \Iltis. We devised teaching material and used it in a large undergraduate course. An important direction for future research is to explore whether our material reduces students' reluctance and initial barriers to engaging with computational reductions in assignments. The observation that our material provides a sense of achievement can serve as  a starting point. It would also be beneficial to assess whether our material is effective for learning algorithmic problems and specific computational reductions. Similarly, exploring the extent to which students can transfer their knowledge of specific reductions and building blocks to the design of other reductions would be a worthwhile direction.

\begin{acks}
\anonymize{This work was supported by the \grantsponsor{DFG}{Deutsche Forschungsgemeinschaft (DFG, German Research Foundation)}{}, grant \grantnum{DFG}{448468041}.}{}
\end{acks}

\printbibliography

\end{document}